\documentclass[11pt,a4paper]{article}

\usepackage{amssymb}

\usepackage[dvips]{graphicx}

\begin{document}

\title{\bf Equation for one-loop divergences in two dimensions and its application to higher spin fields.}

\author{
H.P.Popova\\
{\small{\em Skobeltsyn Institute of Nuclear Physics MSU,}}\\
{\small{\em space science division,}}\\
{\small{\em 119234, Moscow, Russia}},\\
\\
K.V.Stepanyantz\\
{\small{\em Moscow State University,}}\\
{\small{\em physical faculty, department of theoretical physics,}}\\
{\small{\em 119991, Moscow, Russia}}}

\maketitle

\begin{abstract}
A simple formula for one-loop logarithmic divergences on the
background of a two-dimensional curved space-time is derived for
theories for which the second variation of the action is a
nonminimal second order operator with small nonminimal terms. In
particular, this formula allows to calculate terms which are
integrals of total derivatives. As an application of the result,
one-loop divergences for the higher spin fields on the constant
curvature background are obtained in a nonminimal gauge, which
depends on two parameters. By an explicit calculation we
demonstrate that with the considered accuracy the result is
gauge independent. Moreover, the result appeared to be independent
of the spin $s$ for $s\ge 3$.
\end{abstract}

\unitlength=1cm

Keywords: one-loop divergences, higher spin fields.


\section{Introduction}
\hspace{\parindent}

Obtaining one-loop divergences is a typical calculational problem
in quantum field theory. A standard method to solve it is to
construct all divergent diagrams and to calculate them. However,
this can lead to certain difficulties in some cases. For example,
if the calculations are made in the curved space, a number of
divergent diagrams is infinite in the case of constructing the perturbation
theory on the flat background. Nevertheless, writing
the result in the covariant form allows to find a sum of all divergent
contributions in this case. The covariant result can be constructed,
for example, by using the calculations in the weak field limit
in which the deviation of the metric from the flat one is considered
to be small or by some other methods. However, the calculations
can be considerably simplified if some operations are made in the
general case. This is done, for example, within the method for calculating
one-loop divergences proposed by G.t'Hooft and M.Veltman
\cite{'tHooft:1974bx}, by which the one-loop divergences was
first calculated for gravity.

The essence of this method is that if the second variation of the action
(which determines the one-loop divergences) is the minimal second order
operator, then it is possible to construct a covariant equation relating
one-loop divergences to the coefficients of this operator. Then it is not already
necessary to restore a covariant result from the weak field expansion in
each particular case. The t'Hooft--Veltman method allows to
considerably simplify calculations of the one-loop divergences and
to find easily their signs and coefficients. The main shortcoming
of this method is strong limitations on a form of the second
variation of the action. Using various techniques some
generalizations of this method were constructed. In particular,
the t'Hooft--Veltman formula has been generalized to the case of
an arbitrary differential operator \cite{Pronin:1996rv}. Although
the result appeared to be very large, it has been applied for
making some one-loop calculations
\cite{Kazakov:1997nq,Kalmykov:1998cv,Kazakov:1998qt,Kazakov:1999qd,Andriyash:2002yb}.
Moreover, the obtained general formula allows to simplify automatization of the
calculations, if the software for treating tensors (e.g.,
\cite{Bolotin:2013qgr,Poslavsky:2015vba}) is used.

From the mathematical point of view the t'Hooft--Veltman method
corresponds to calculating the Minakshisundaram--Seeley--de-Witt coefficient
\cite{Minakshisundaram:1949xg,Minakshisundaram:1953xh,Seeley:1967ea,Gilkey:1975iq}
$b_4$ for a differential operator coinciding with the second
variation of the action. In four dimensions this coefficient is
related to one-loop logarithmic divergences. In the general case, in the
space of the dimension $d=2n$ the one-loop logarithmic divergences
are related to the coefficient $b_{2n}$.

For various differential operators
the Minakshisundaram--Seeley--de-Witt coefficients can be calculated
using the Schwinger--de-Witt technique and its generalizations
\cite{Schwinger:1951nm,DeWitt:1965jb,Barvinsky:1985an,Gusynin:1990ek}.
However, this technique does not allow to obtain a result for an arbitrary differential
operator. From the other side, the total derivative terms were not
found in Ref. \cite{Pronin:1996rv}. (Their contributions to the
one-loop divergences are integrals of total derivatives.) Therefore,
there is a problem how the results of Ref. \cite{Pronin:1996rv} can
be generalized in order to take into account total derivative terms.
These contributions are essential in some cases, for example, if the
calculations are made on the (anti) de-Sitter background. Let us note
that the calculations on the (anti) de-Sitter background are usually
made by different methods. For example, the Minakshisundaram--Seeley--de-Witt
coefficients for the minimal operator can be found in an arbitrary
dimension by the harmonic analysis on the homogeneous spaces
\cite{Camporesi:1990wm}. Using this method one-loop divergences were
calculated for the fields of an arbitrary spin on the four-dimensional
(anti) de-Sitter background in the minimal gauge \cite{Camporesi:1993mz}.
A similar calculation, based on the formula for the $b_4$ coefficient
of the minimal operator, was made in Ref. \cite{Takata}.

In the present paper we try to understand, how it is possible to
calculate total derivatives terms using the t'Hooft--Veltman
technique for nonminimal operators. For this purpose we consider
the simplest case of the two-dimensional space, in which logarithmic divergences are
related to the coefficient $b_2$. We calculate this coefficient for the second order
nonminimal operator taking into consideration the total derivative terms,
under the assumption that nonminimal terms are small, but
nonvanishing. Then this result is verified by a calculation of
one-loop divergences for the higher spin theory in a nonminimal
gauge, which depends on two arbitrary (small) parameters.

The paper is organized as follows. In Sect. \ref{Section_B2} we
describe a method for calculating one-loop divergences which is
applicable, if the second variation of the action is a second
order nonminimal operator with small nonminimal terms, and present
the result of this calculation, which is given by Eq. (\ref{B2}).
Using this result in Sect. \ref{Section_One_Loop} a divergent part
of the one-loop effective action for the higher spin theory on the
two-dimensional (anti) de-Sitter background is calculated in a
nonminimal gauge depending on two parameters. For this purpose in
Sect. \ref{Subsection_Higher_Spins} we recall the basic
information about the higher spin theory on the constant curvature
background. The calculation of one-loop divergences is described
in Sect. \ref{Subsection_Calculation}. The result is given by Eq.
(\ref{Result}). The results are briefly discussed in the
conclusion. One technical problem related to the one-loop
calculation is considered in the Appendix.

\section{One-loop divergences in two dimensions for second order
nonminimal operator}
\hspace{\parindent}\label{Section_B2}

In the one-loop approximation the effective action is written in the form
(see, e.g., \cite{Huang:1982ik})

\begin{equation} \Gamma[\varphi]=S[\varphi]+ \frac {i}{2}\hbar\,
\mbox{Tr}\,  \ln D + O(\hbar^2),
\end{equation}

\noindent where the operation $\mbox{Tr}$ by definition includes
$\int d^dx$, and the differential operator $D$ is the second variation
of the classical action:

\begin{equation}
D_i{}^j = \frac{\delta^2 S}{\delta\varphi^i \delta\varphi_j},
\end{equation}

\noindent where each of the indexes $i$ and $j$ denotes the whole
set of possible indexes of the fields $\varphi$. Most models considered in
field theory are quadratic in the derivatives of fields.
Therefore, for a large number of practical problems it is
sufficient to consider only differential operators of the second
order. Note that in some cases differential operators of
higher orders are also interesting, see, for example, Ref.
\cite{Pronin:1997eb}. An arbitrary nonminimal differential
operator of the second order has the form

\begin{equation}
D_i{}^j = (I g^{\mu\nu} \nabla_\mu \nabla_\nu + K^{\mu\nu}
\nabla_\mu \nabla_\nu + S^\mu \nabla_\mu + W)_i{}^j,
\end{equation}

\noindent where for later convenience we extract the terms
containing the Laplace operator, and the covariant derivative is
written as

\begin{equation}
(\nabla_\mu)_i{}^j = \delta_i^j \partial_\mu + \omega_\mu{}_i{}^j,
\end{equation}

\noindent where $\omega_\mu{}_i{}^j$ is a connection. The
coefficients $I_i{}^j$, $K^{\mu\nu}{}_i{}^j$, $S^\mu{}_i{}^j$, and
$W_i{}^j$ are functions of the fields and are obtained by
calculating the second variation of the action. Without loss of
generality $K^{\mu\nu}{}_i{}^j$ can be assumed to be symmetric in
the indexes $\mu$ and $\nu$ that will be always assumed below. The
term ``nonminimal'' means that terms with a maximal number of the
derivatives do not coincide with the Laplace operator in a certain
degree. (In the considered case these are the terms with the
second derivatives, which should be compared with the first degree
of the Laplace operator.)

We will calculate one-loop divergences using the dimensional
regularization
\cite{'tHooft:1972fi,Bollini:1972ui,Ashmore:1972uj,Cicuta:1972jf}.
The space-time dimension we will denote by $d$. If a theory is
considered in two dimensions, then the divergent part of the
one-loop effective action is proportional to $(d-2)^{-1}$, where
$d\to 2$ after the renormalization. Using the general coordinate invariance, it is
convenient to present the divergent part of the one-loop effective
action in the form

\begin{equation}
\Gamma_{1-loop}^{(\infty)} = \frac{1}{4\pi(d-2)}\int d^2x
\sqrt{-g}\, b_2,
\end{equation}

\noindent where $b_2$ is a covariant function of the fields. This
function is the second Minakshisundaram--Seeley--de-Witt
coefficient. In this paper we calculate it for the second order nonminimal
differential operator, assuming that the nonminimal terms are small. Moreover,
we will assume that $(S^\mu)_i{}^j = 0$ that is valid for
a large number of practical problems.

In order to find one-loop divergences we will use the
generalization of the method proposed by G.t'Hooft and M.Veltman
\cite{'tHooft:1974bx}, using which the coefficient $b_4$ was
calculated for an arbitrary differential operator without taking
into account the total derivative terms \cite{Pronin:1996rv}. In
this paper we will not already omit the total derivative terms.
We consider the operator

\begin{equation}\label{Operator_S}
D_i{}^j = (I g^{\mu\nu} \nabla_{\mu} \nabla_\nu +\varepsilon
K^{\mu\nu} \nabla_{\mu}\nabla_{\nu} + W)_i{}^j,
\end{equation}

\noindent assuming that $\varepsilon \to 0$ is a small parameter.
We will calculate one-loop divergences up to the terms of the first
order in $\varepsilon$. Without loss of generality it is possible to
consider (as we do) that the matrix $\varepsilon
K^{\mu\nu}{}_i{}^j$ is symmetric in the indexes $\mu$ and $\nu$.
Moreover, we also assume that this matrix and the matrix $I_i{}^j$
depend only on the metric tensor $g_{\alpha\beta}$ (certainly,
$\delta_\alpha^\beta$ and $g^{\alpha\beta}$ are also possible).

For constructing one-loop divergences we will use the diagram
technique. In order to take into account total derivative terms within this
technique, we multiply the logarithm of the operator $D$ by an
auxiliary function $a(x)$ which has not any indexes and
calculate a trace of the result:

\begin{equation}\label{We_Calculate}
\Big(\mbox{Tr}\, a(x) \ln D\Big)^{(\infty)} = \Big(\int d^2x\, a(x)\, \mbox{tr} \ln D\Big)^{(\infty)} =
\frac{2}{i}\cdot \frac{1}{4\pi(d-2)} \int d^2x\,\sqrt{-g}\,a(x)\,
b_2,
\end{equation}

\noindent where $\mbox{tr}$ denotes the usual matrix trace with
respect to the indexes $i$ and $j$. Setting $a(x)=1$ we obtain the
one-loop contribution to effective action up to the factor $i/2$.
However, if it is necessary to take carefully into account terms
which are integrals of total derivatives, then the presence of a
nontrivial function $a(x)$ is needed. The expression (\ref{We_Calculate})
can be presented as a sum of one-loop diagrams in which one of external lines
corresponds to the function $a(x)$. For this purpose it is
convenient to rewrite the operator $D$ in the following form:

\begin{equation}
D_i{}^j \equiv I_0{}_i{}^j {\partial}^{2}_{\mu}+\varepsilon
K_0^{\mu\nu}{}_i{}^j \partial_{\mu}\partial_{\nu} +V_i{}^j,
\end{equation}

\noindent where the matrices $I_0$ and $K_0^{\mu\nu}$ are obtained
from the matrices $I$ and $K^{\mu\nu}$ by substituting the metric
tensor $g_{\mu\nu}$ (or $g^{\mu\nu}$) by the flat space metric
$\eta_{\mu\nu}$ (or $\eta^{\mu\nu}$). The operator $V$ includes
all other terms. Taking into account the equality

\begin{equation}
\ln(I_0 \partial_\mu^2) = \ln(I_0) + \ln(\partial_\mu^2),
\end{equation}

\noindent diagrams corresponding to the expression
(\ref{We_Calculate}) can be constructed using the expansion

\begin{eqnarray}\label{Expansion}
&&\hspace*{-5mm} \ln(D) =\ln(I_0
\partial^2_{\mu})
+\ln\Big(1+\frac{1}{\partial_\mu^2} I_0^{-1} (\varepsilon
K_0^{\mu\nu}\partial_\mu \partial_\nu + V) \Big)
\nonumber\\
&&\hspace*{-5mm} +\frac{1}{2}\Big[\ln(I_0) + \ln
(\partial^2_{\mu}),\ln\Big(1+\frac{1}{\partial_\mu^2} I_0^{-1}
(\varepsilon K_0^{\mu\nu}\partial_\mu \partial_\nu + V)
\Big)\Big]\nonumber\\
&&\hspace*{-5mm} +\frac{1}{12}\Big[\ln(I_0) + \ln
(\partial^2_{\mu}),\Big[\ln(I_0) + \ln (\partial^2_{\mu}),
\ln\Big(1+\frac{1}{\partial_\mu^2} I_0^{-1} (\varepsilon
K_0^{\mu\nu}\partial_\mu \partial_\nu + V) \Big)\Big]\Big]
\nonumber\\
&&\hspace*{-5mm}
+ \frac{1}{12}\Big[\ln\Big(1+\frac{1}{\partial_\mu^2} I_0^{-1} (\varepsilon
K_0^{\mu\nu}\partial_\mu \partial_\nu + V) \Big),\Big[\ln\Big(1+\frac{1}{\partial_\mu^2} I_0^{-1} (\varepsilon
K_0^{\mu\nu}\partial_\mu \partial_\nu + V) \Big),\nonumber\\
&&\hspace*{-5mm} \ln(I_0) + \ln (\partial^2_{\mu})\Big]\Big]+
\ldots,\vphantom{\frac{1}{2}}
\end{eqnarray}

\noindent
where dots denote terms with a larger number of commutators.

The first term in this expression does not depend on
fields. Its contribution to the one-loop divergences is an
insignificant constant. The other terms contain

\begin{equation}
\ln\Big(1+\frac{1}{\partial_\mu^2} I_0^{-1} (\varepsilon
K_0^{\mu\nu}\partial_\mu \partial_\nu + V) \Big),
\end{equation}

\noindent which can be easily expanded in powers of $V$ and
$\varepsilon$. All terms of the zero and the first order in
$V$ or $\varepsilon$ containing a commutator with $\ln(I_0)$
vanish after calculating the matrix trace, because the function
$a(x)$ has no matrix indexes.

\begin{figure}[h]
\begin{picture}(0,2.4)
\put(0.3,1.6){$(1)$} \put(1.1,1.4){$W$} \put(1.2,0.4){$a$}
\put(1.5,0.4){\includegraphics[scale=0.3]{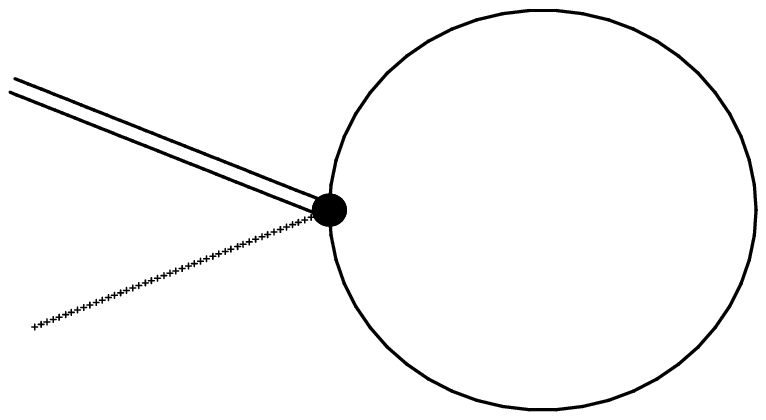}}
\put(4.3,1.6){$(2)$} \put(5.0,1.4){$\omega_\mu$}
\put(5.2,0.4){$a$}
\put(5.5,0.4){\includegraphics[scale=0.3]{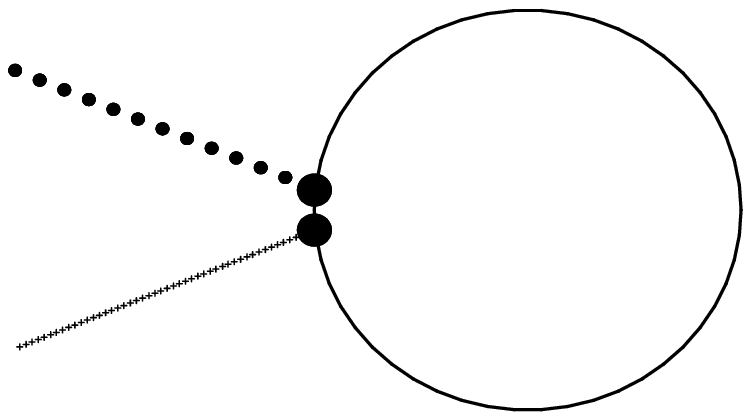}}
\put(8.3,1.6){$(3)$} \put(9.0,1.4){$h_{\mu\nu}$}
\put(9.2,0.4){$a$}
\put(9.5,0.4){\includegraphics[scale=0.3]{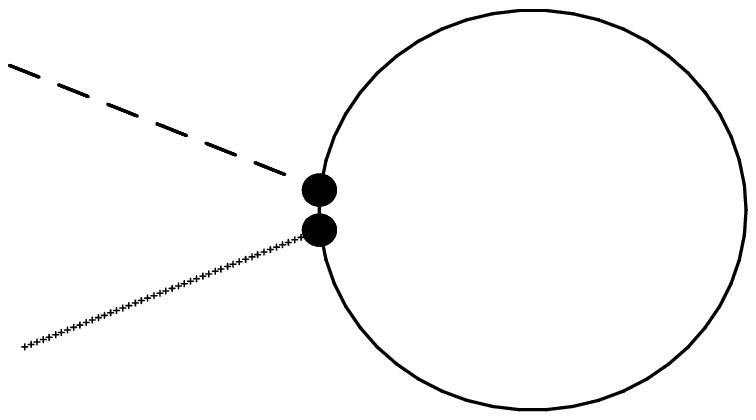}}
\put(12.3,1.6){$(4)$} \put(13.1,1.4){$\phi$} \put(13.2,0.4){$a$}
\put(13.5,0.4){\includegraphics[scale=0.3]{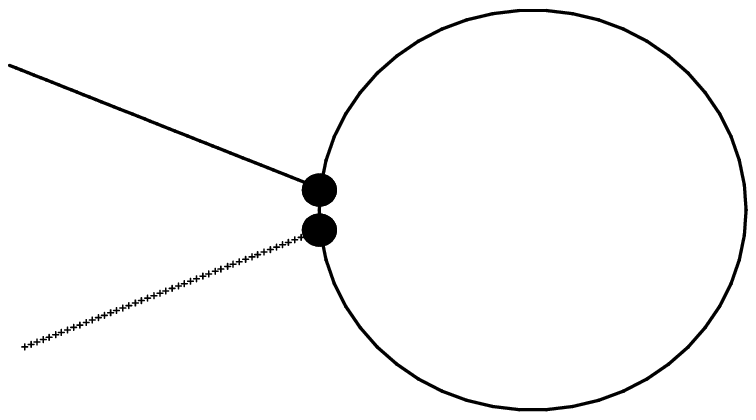}}
\end{picture}
\caption{Diagrams in $d=2$, giving the divergent part of the
one-loop effective action.} \label{Figure_Diagrams}
\end{figure}

The remaining terms give the sum of Feynman diagrams in the one-loop approximation.
As we already mentioned, one of the external lines in these diagrams corresponds to the
function $a(x)$. We extract from these diagrams the divergent ones.
If the calculations are made on the flat background, then there is the
only divergent diagram (1) in Fig. \ref{Figure_Diagrams}.

Using Eqs. (\ref{We_Calculate}) and (\ref{Expansion}) we construct analytical
expressions for these diagrams, from which we extract logarithmically divergent terms.
As usual, they are calculated in the Euclidean space after the Wick rotation.
In the considered terms we integrate over angles and, after that, replace the remaining
Euclidean integral over the momentum according to the prescription

\begin{equation}
\int \frac{d^dk}{(2\pi)^2 k^2} \to -\frac{1}{2\pi(d-2)}.
\end{equation}

In order to generalize the result to the case of the curved
space-time it is convenient to use the weak field expansion
around the metric of the flat space-time. For this purpose we
define

\begin{equation}
h_{\mu\nu} \equiv g_{\mu\nu} - \eta_{\mu\nu},
\end{equation}

\noindent which is a deviation of the metric from the flat metric
$\eta_{\mu\nu}$. We assume that this expression is small. A number of divergent
diagrams in which external lines correspond to $h_{\mu\nu}$ is infinite.
However, a sum of these diagrams is a weak field expansion of
a certain covariant result. From dimensional arguments it is easy to see
that terms which can appear in the calculation of one-loop divergences
should contain the curvature in no more than the first degree.
Therefore, from the infinite set of the divergent diagrams it is sufficient
to calculate only the diagrams $(2)$ --- $(4)$ in Fig. \ref{Figure_Diagrams}.
Two small adjacent circles mean that, for example, the derivatives $\partial_\mu$
can appear between $h_{\mu\nu}$ and $a$. $\phi^a$ (as in Ref.
\cite{Pronin:1996rv}) denotes the fields $h_{\alpha\beta}$ which
arise from the expansion of $\varepsilon K^{\mu\nu}{}_i{}^j$ in
powers of the deviation of the metric from the flat one. These
fields are excluded from the final result by using the identity

\begin{equation}
0 = \nabla_\alpha K^{\mu\nu}{}_i{}^j = \frac{\partial}{\partial
\phi^a} K^{\mu\nu}{}_i{}^j \partial_\alpha\phi^a +
\Gamma_{\alpha\beta}^\mu K^{\beta\nu}{}_i{}^j +
\Gamma_{\alpha\beta}^\nu K^{\mu\beta}{}_i{}^j +
\omega_\alpha{}_i{}^k K^{\mu\nu}{}_k{}^j - K^{\mu\nu}{}_i{}^k
\omega_\alpha{}_k{}^j.
\end{equation}

Calculating the diagrams $(1)$ --- $(4)$ according to the
algorithm described above and constructing the covariant result
using the equations

\begin{eqnarray}
&& R_{\mu\nu} = \frac{1}{2}\Big(\partial_\mu\partial_\alpha
h_{\nu}{}^{\alpha} + \partial_\nu\partial_\alpha h_{\mu}{}^{\alpha} -
\partial_\mu\partial_\nu h_\alpha{}^\alpha - \partial^2 h_{\mu\nu} \Big) +O(h^2);\nonumber\\
&& R = \partial_\mu\partial_\nu h^{\mu\nu} - \partial^2 h_\alpha{}^\alpha +
O(h^2);\nonumber\\
&& F_{\mu\nu}{}_i{}^j = \partial_\mu \omega_\nu{}_i{}^j - \partial_\nu \omega_\mu{}_i{}^j + O(\omega^2),\vphantom{\frac{1}{2}}
\end{eqnarray}

\noindent finally we obtain

\begin{equation}\label{B2}
b_{2} = \mbox{tr} \Big(\widehat W +\frac{1}{6}R
-\frac{1}{2}\varepsilon \widehat K_{\alpha}{}^{\alpha}\widehat W
-\frac{1}{12}\varepsilon \widehat K_{\alpha}{}^{\alpha}
R+\frac{1}{6}\varepsilon \widehat K^{\mu\nu} R_{\mu\nu}\Big),
\end{equation}

\noindent where we use the notation

\begin{equation}
\varepsilon \widehat K^{\mu\nu} \equiv I^{-1} \varepsilon
K^{\mu\nu}; \qquad \widehat W \equiv I^{-1} W.
\end{equation}

\noindent The possibility of writing the result in the covariant
form can be considered as a nontrivial test of the calculations correctness.
Also we note that the result is written in a very compact form (in
comparisons with the expression for one-loop divergences in the
case of an arbitrary nonminimal operator). Therefore, it
is possible to suggest that in the limit $\varepsilon \to 0$ the
formula for one-loop divergences in $d=4$ can be also considerably
simplified.

For $\varepsilon K^{\mu\nu}{}_i{}^j = \varepsilon g^{\mu\nu}
\delta_i^j$ and $I=1$ the formula (\ref{B2}) correctly reproduces
the known result for the $b_2$ coefficient of the minimal operator
(see, for example, \cite{Obukhov:1983mm}) including the total
derivative terms.

\section{One-loop divergences for higher spin fields in the
nonminimal gauge}\label{Section_One_Loop}

\subsection{Higher spin fields on the (A)dS background}
\hspace{\parindent}\label{Subsection_Higher_Spins}

Higher spins are described by the totally symmetric tensor fields
$\phi_{\mu_1\mu_2 \ldots \mu_s}$ which satisfy the double
tracelessness condition

\begin{equation}
\phi_{\alpha}{}^\alpha{}_\beta{}^\beta{}_{\mu_5\ldots\mu_s} = 0.
\end{equation}

\noindent
(In this paper we will consider the case $s\ge 3$.) The free action for
these fields on background of the flat space-time has been constructed in Ref.
\cite{Fronsdal:1978rb}. (Fermion higher spin fields and the action
for them are described in Ref. \cite{Fang:1978wz}.) It is possible
to construct an interacting theory for the higher spins
\cite{Vasiliev:2004qz,Bekaert:2005vh}, but this is a very complicated
problem. Even an action quadratic in the higher spin fields on the curved
background \cite{Fronsdal:1978vb} can be written only if the background
geometry is a constant curvature space ((anti) de-Sitter space),
for which

\begin{equation}
R_{\mu\nu\alpha\beta}=\frac{1}{d(d-1)}(g_{\mu\alpha}g_{\nu\beta}-g_{\mu\beta}g_{\nu\alpha})R,
\end{equation}

\noindent where $R =\mbox{const}$. It was shown in Ref.
\cite{Buchbinder:2011vp} that under certain assumptions a
consistent Lagrangian formulation for the free boson totally
symmetric higher spin fields is possible only in this case. The
action for the higher spin fields on the (anti) de-Sitter
background is written as

\begin{eqnarray}\label{Fronsdal_Action}
&& S=\frac{(-1)^s}{2}\int d^d x\sqrt{-g}
\Big[(\nabla_{\alpha}\phi_{\mu_1\ldots
\mu_s})^{2}-\frac{1}{2}s(s-1)
(\nabla_{\alpha}\phi_{\beta}{}^\beta{}_{\mu_3\ldots \mu_s})^2
-s(\nabla^{\alpha}\phi_{\alpha \mu_2\ldots
\mu_s})^2\nonumber\\
&& +s(s-1) \nabla_{\alpha}\phi^{\alpha\beta\mu_3\ldots
\mu_s}\nabla_{\beta}\phi_{\gamma}{}^\gamma{}_{\mu_3\ldots \mu_s}
-\frac{1}{4}s(s-1)(s-2)
(\nabla_{\alpha}\phi^{\alpha\beta}{}_{\beta\mu_4\ldots
\mu_s})^2 + c_1 R (\phi_{\mu_1\ldots \mu_s})^{2} \nonumber\\
&&  + c_2 R (\phi_{\gamma}{}^\gamma{}_{\mu_3\ldots \mu_s})^2 \Big],
\end{eqnarray}

\noindent where the squares of tensors denote contractions of all
free indexes with the metric $g^{\mu\nu}$, and the coefficients $c_1$ and $c_2$ are

\begin{equation}
c_1 = - \frac{(s-1)(s-4)}{d(d-1)} - \frac{(s-2)}{d}; \qquad c_2 =
\frac{s(s-1)}{2d} \Big(s-1 + \frac{s(s-3)}{d-1}\Big).
\end{equation}

\noindent These values are obtained by requiring the gauge invariance
of the action under the transformations

\begin{equation}
\delta\phi_{\mu_1\ldots\mu_s}
=\frac{1}{s}(\nabla_{\mu_1}\alpha_{\mu_2\ldots\mu_s}
+\nabla_{\mu_2}\alpha_{\mu_1\mu_3\ldots\mu_s}+\ldots).
\end{equation}

\noindent Their parameter $\alpha_{\mu_1\mu_2\ldots \mu_{s-1}}$ is
a totally symmetric tensor which satisfies the tracelessness condition
$\alpha_\beta{}^\beta{}_{\mu_3\ldots\mu_{s-1}} =
0.$ In a particular case $d=2$, which is considered in this paper, the coefficients
$c_1$ and $c_2$ are given by

\begin{equation}
c_1 = -\frac{1}{2} (s^2-4s+2); \qquad c_2 = \frac{s(s-1)}{4}
(s^2-2s-1).
\end{equation}

For quantization of gauge theories \cite{Faddeev:1980be} (and, in
particular, the considered higher spin theory on the (anti)
de-Sitter background) it is necessary to fix a gauge and add the
corresponding Faddeev--Popov ghosts \cite{Faddeev:1967fc}. It is
well known that the effective action is gauge independent on
shell. However, explicit calculations (see, e.g., Ref.
\cite{Kallosh:1978wt}) show that the effective action depends on a
gauge off shell. In this paper a gauge is fixed by adding the
terms

\begin{eqnarray}\label{gauge2}
&& S_{\mbox{\scriptsize gf}} =\frac{(-1)^s}{2}\int
d^dx\,\sqrt{-g}\,
s(1+\lambda)\Big(\nabla^{\alpha}\phi_{\alpha\mu_1\ldots \mu_{s-1}}
-\frac{1}{2}(s-1)(1+\beta)\nabla_{(\mu_1}
\phi^{\alpha}{}_{\alpha\mu_2\ldots \mu_{s-1})}\nonumber\\
&& + \frac{\beta (s-1)(s-2)}{2d + 4(s-3)}\, g_{(\mu_1\mu_2}
\nabla_\alpha \phi^{\alpha\beta}{}_{\beta\mu_3\ldots \mu_{s-1})}
\Big)^2.
\end{eqnarray}

\noindent The last term is added in order that the gauge
condition is traceless, and a number of the gauge conditions
coincides with a number of gauge transformation parameters. The
round brackets denote symmetrization with respect to the indexes
$\mu_1\mu_2\ldots\mu_{s-1}$. In the general case,

\begin{equation}
T_{(\mu_1\mu_2\ldots\mu_k)} \equiv
\frac{1}{k!}\Big(T_{\mu_1\mu_2\ldots\mu_k} +
T_{\mu_2\mu_1\ldots\mu_k} + \mbox{the other permutations of indexes}\Big).
\end{equation}

The Lagrangian for the Faddeev--Popov ghosts is obtained in a
standard way by making a gauge transformation in the expression
for the gauge condition

\begin{eqnarray}
&& \nabla^{\alpha}\phi_{\alpha\mu_1\ldots \mu_{s-1}}
-\frac{1}{2}(s-1)(1+\beta)\nabla_{(\mu_1}
\phi^{\alpha}{}_{\alpha\mu_2\ldots \mu_{s-1})}\nonumber\\
&& \qquad\qquad\qquad\qquad\qquad\qquad\quad + \frac{\beta
(s-1)(s-2)}{2d + 4(s-3)}\, g_{(\mu_1\mu_2} \nabla_\alpha
\phi^{\alpha\beta}{}_{\beta\mu_3\ldots \mu_{s-1})}.\qquad
\end{eqnarray}

\noindent The gauge parameter becomes the ghost field
$c_{\mu_1\mu_2\ldots\mu_{s-1}}$, and the result is multiplied by
the antighost field $\overline{c}^{\mu_1\mu_2\ldots\mu_{s-1}}$.
As well as the parameters of the gauge transformation, the ghost
and antighost fields are totally symmetric and traceless. It is easy to
see that the result for the ghost Lagrangian is written as

\begin{eqnarray}
&& L_{\mbox{\scriptsize
gh}}=\overline{c}^{\mu_1\mu_2\ldots\mu_{s-1}}\Big(\nabla^\alpha \nabla_{\alpha}
c_{\mu_1\mu_2\ldots\mu_{s-1}}-\frac{\beta}{2}(s-1)(\nabla_{\mu_1}\nabla^{\alpha}
+\nabla^{\alpha}\nabla_{\mu_1})c_{\alpha\mu_2\ldots\mu_{s-1}}
\nonumber\\
&& + (\beta+2)\frac{(s-1)(d+s-3)}{2 d(d-1)} R c_{\mu_1\mu_2\ldots
\mu_{s-1}}\Big).\qquad
\end{eqnarray}

\subsection{Calculation of one-loop divergences}
\hspace{\parindent}\label{Subsection_Calculation}

In this paper we use Eq. (\ref{B2}) for calculating a divergent
part of the one-loop effective action for the considered theory,
if $d=2$, $s\ge 3$, and the parameters $\lambda$ and $\beta$ are small. In
this case in the lowest order in $\lambda$ and $\beta$ the sum of
the classical action and the gauge fixing term can be written in
the form

\begin{eqnarray}\label{S+S_gh}
&& S+S_{\mbox{\scriptsize gf}}=\frac{(-1)^s}{2}\int d^2 x
\sqrt{-g}\Big\{(\nabla_{\alpha}\phi_{\mu_1\mu_2\ldots\mu_s})^{2}
-\frac{1}{4}s(s-1)(1 -\lambda -2\beta)
(\nabla_{\alpha}\phi^{\beta}{}_{\beta \mu_1\ldots\mu_{s-2}})^2\nonumber\\
&& +\frac{\lambda s}{2}\Big( (\nabla^{\alpha}\phi_{\alpha
\mu_1\ldots\mu_{s-1}})^2 + \nabla_\alpha
\phi_{\beta\mu_1\ldots\mu_{s-1}} \nabla^\beta
\phi^{\alpha\mu_1\ldots\mu_{s-1}}\Big) -s(s-1)(\lambda
+\beta)\nabla^{\alpha}\phi_{\alpha\beta \mu_1\ldots\mu_{s-2}}
\nonumber\\
&& \times \nabla^{\beta}\phi_{\gamma}{}^{\gamma
\mu_1\ldots\mu_{s-2}} +\frac{1}{8}s(s-1)(s-2)(\lambda+2\beta)
\Big((\nabla_{\alpha}\phi^{\alpha\beta}{}_{\beta\mu_1\ldots
\mu_{s-3}})^2 +
\nabla_\gamma\phi^{\alpha\beta}{}_{\beta\mu_1\ldots\mu_{s-3}}
\nonumber\\
&& \times
\nabla_\alpha\phi_\delta{}^{\delta\gamma\mu_1\ldots\mu_{s-3}}
\Big) +R (\phi_{\mu_1 \mu_2\ldots
\mu_s})^{2}\Big(-\frac{1}{2}(s^2-4s+2) + \frac{1}{4} \lambda s^2
\Big) + \frac{1}{4} R (\phi^{\alpha}{}_{\alpha
\mu_1\ldots\mu_{s-2}})^2
\nonumber\\
&& \times s(s-1) \Big(s^2-2s-1 - \lambda - \frac{1}{4} (s-2)^2 (2
+ \lambda + 2 \beta) \Big)  + o(\lambda,\beta)\Big\},
\end{eqnarray}

\noindent
where $o(\lambda,\beta)$ denotes terms of higher orders in $\lambda$ and $\beta$.

The second variation of this expression with respect to
the fields $\phi_{\alpha_1\alpha_2\ldots\alpha_s}$ is a second
order differential operator. Calculating a trace of the logarithm
of this operator, we obtain one-loop diagrams with a loop of the
spin $s$ field and external lines corresponding to the field
$h_{\mu\nu}$. However, it is not necessary to calculate the
diagrams in this case, because one can use Eq. (\ref{B2}), which
immediately gives the sum of their divergent parts in the covariant form.
In order to use this formula, it is necessary to find the second
variation of Eq. (\ref{S+S_gh}) and, using it, obtain the matrices $I$,
$\varepsilon K^{\mu\nu}$, and $W$. Constructing these matrices it
is important to take into account that the fields
$\phi_{\alpha_1\alpha_2\ldots\alpha_s}$ are double traceless.
As a consequence, the projection operators to the double traceless
states appear in all matrices. We will denote them by
$Q_{\alpha_1\alpha_2\ldots\alpha_s}{}^{\beta_1\beta_2\ldots\beta_s}$.
The structure of this projection operator and its properties are
discussed in the Appendix.

Having calculated the second variation of the action we find that
it is given by a differential operator of the form
(\ref{Operator_S}), in which (after omitting an inessential factor)

\begin{eqnarray}\label{Matrixes_Main1}
&& I_{\alpha_1\alpha_2\ldots\alpha_s}
{}^{\beta_1\beta_2\ldots\beta_s} =
Q_{\alpha_1\alpha_2\ldots\alpha_s}
{}^{\gamma_1\gamma_2\ldots\gamma_s} \Big(
1_{\gamma_1\gamma_2\ldots\gamma_s} ^{\delta_1\delta_2\ldots\delta_s}
- \frac{s(s-1)}{4} \cdot g_{(\gamma_1\gamma_2} g^{(\delta_1\delta_2}
1_{\gamma_3\ldots\gamma_s)}^{\delta_3\ldots\delta_s)}\Big)\nonumber\\
&&\times Q_{\delta_1\delta_2\ldots\delta_s}
{}^{\beta_1\beta_2\ldots\beta_s};\vphantom{\frac{1}{2}}\qquad\\
&& \nonumber\\
\label{Matrixes_Main2}
&& \varepsilon K^{\mu\nu}{}_{\alpha_1\alpha_2\ldots\alpha_s}
{}^{\beta_1\beta_2\ldots\beta_s} =
Q_{\alpha_1\alpha_2\ldots\alpha_s}
{}^{\gamma_1\gamma_2\ldots\gamma_s} \Big\{
\frac{s(s-1)}{4}(\lambda+2\beta) g^{\mu\nu} g_{(\gamma_1\gamma_2}
g^{(\delta_1\delta_2}
1_{\gamma_3\ldots\gamma_s)}^{\delta_3\ldots\delta_s)}\nonumber\\
&& + \frac{s \lambda}{2} \Big(\delta^\mu_{(\gamma_1}
g^{\nu(\delta_1}
1_{\gamma_2\ldots\gamma_s)}^{\delta_2\ldots\delta_s)} +
\delta^\nu_{(\gamma_1} g^{\mu(\delta_1}
1_{\gamma_2\ldots\gamma_s)}^{\delta_2\ldots\delta_s)} \Big)
\nonumber\\
&& - \frac{s(s-1)}{2}(\lambda+\beta) \Big(
\delta^{\mu}_{(\gamma_1} \delta^{\nu}_{\gamma_2}
g^{(\delta_1\delta_2}
1_{\gamma_3\ldots\gamma_s)}^{\delta_3\ldots\delta_s)} +
g^{\mu(\delta_1} g^{\nu\delta_2} g_{(\gamma_1\gamma_2}
1_{\gamma_3\ldots\gamma_s)}^{\delta_3\ldots\delta_s)} \Big)
\nonumber\\
&& + \frac{s(s-1)(s-2)}{8} (\lambda+2\beta)
\Big(\delta^\mu_{(\gamma_1} g^{\nu(\delta_1} g^{\delta_2\delta_3}
g_{\gamma_2\gamma_3}
1_{\gamma_4\ldots\gamma_s)}^{\delta_4\ldots\delta_s)} +
\delta^\nu_{(\gamma_1} g^{\mu(\delta_1} g^{\delta_2\delta_3}
g_{\gamma_2\gamma_3}
1_{\gamma_4\ldots\gamma_s)}^{\delta_4\ldots\delta_s)}\Big)
\Big\}\nonumber\\
&& \times Q_{\delta_1\delta_2\ldots\delta_s}
{}^{\beta_1\beta_2\ldots\beta_s};\vphantom{\frac{1}{2}}\\
&&\nonumber\\
\label{Matrixes_Main3}
&& W_{\alpha_1\alpha_2\ldots\alpha_s}
{}^{\beta_1\beta_2\ldots\beta_s} =
Q_{\alpha_1\alpha_2\ldots\alpha_s}
{}^{\gamma_1\gamma_2\ldots\gamma_s} \Big\{ R
\Big(\frac{1}{2}(s^2-4s+2) - \frac{1}{4} \lambda s^2
\Big)\cdot 1_{\gamma_1\gamma_2\ldots\gamma_s}^{\delta_1\delta_2\ldots\delta_s}\nonumber\\
&& - \frac{1}{4} R s(s-1) \Big(s^2-2s-1 - \lambda - \frac{1}{4}
(s-2)^2 (2 + \lambda + 2 \beta) \Big)\cdot g_{(\gamma_1\gamma_2}
g^{(\delta_1\delta_2}
1_{\gamma_3\ldots\gamma_s)}^{\delta_3\ldots\delta_s)} \Big\}\nonumber\\
&& \times Q_{\delta_1\delta_2\ldots\delta_s}
{}^{\beta_1\beta_2\ldots\beta_s},\vphantom{\frac{1}{2}}
\end{eqnarray}

\noindent where we use the notation

\begin{equation}
1_{\gamma_1\gamma_2\ldots\gamma_k} ^{\beta_1\beta_2\ldots\beta_k}
\equiv \delta^{(\beta_1}_{(\gamma_1} \delta^{\beta_2}_{\gamma_2}
\ldots \delta^{\beta_k)}_{\gamma_k)}.
\end{equation}

\noindent The matrix inverse to $I$ is defined by

\begin{equation}
I_{\alpha_1\alpha_2\ldots\alpha_s}
{}^{\beta_1\beta_2\ldots\beta_s}
(I^{-1})_{\beta_1\beta_2\ldots\beta_s}
{}^{\gamma_1\gamma_2\ldots\gamma_s} =
Q_{\alpha_1\alpha_2\ldots\alpha_s}{}^{\gamma_1\gamma_2\ldots\gamma_s}.
\end{equation}

\noindent From this condition one can obtain an explicit
expression for the matrix $I^{-1}$, which has the form

\begin{equation}\label{I^(-1)}
(I^{-1})_{\alpha_1\alpha_2\ldots\alpha_s}
{}^{\beta_1\beta_2\ldots\beta_s} =
Q_{\alpha_1\alpha_2\ldots\alpha_s}
{}^{\gamma_1\gamma_2\ldots\gamma_s}\Big(
1_{\gamma_1\gamma_2\ldots\gamma_s}^{\beta_1\beta_2\ldots\beta_s} - \frac{s(s-1)}{4(s-2)}\cdot
g_{(\gamma_1\gamma_2} g^{(\beta_1\beta_2}
1_{\gamma_3\ldots\gamma_s)}^{\beta_3\ldots\beta_s)}\Big).
\end{equation}

\noindent Substituting the matrices (\ref{Matrixes_Main1}) --- (\ref{Matrixes_Main3}) and
(\ref{I^(-1)}) into Eq. (\ref{B2}) we obtained the following result
for the coefficient $b_2$ corresponding to the second variation of
the action (\ref{S+S_gh}):

\begin{equation}\label{Main}
b_{2\mbox{\scriptsize (main)}}=\Big(2(s-1)^2-\frac{4}{3}+\beta
s(s-1)^2+o(\lambda,\beta)\Big)R.
\end{equation}

\noindent The integral of this expression multiplied by $\sqrt{-g}$ over $d^2x$ is
proportional to the divergent part of the sum of one-loop Feynman
diagrams containing a loop of the spin $s$ field.

The divergent part of the one-loop effective action is also
contributed by diagrams with a loop of the Faddeev--Popov ghosts.
This contribution can be also calculated using Eq. (\ref{B2}),
which should be applied to the second variation of the ghost
action. For $d=2$ the corresponding Lagrangian has the form

\begin{eqnarray}\label{Ghost_Lagrangian}
&& L_{\mbox{\scriptsize
gh}}=\overline{c}^{\mu_1\mu_2\ldots\mu_{s-1}}\Big(\nabla^\alpha \nabla_{\alpha}
c_{\mu_1\mu_2\ldots\mu_{s-1}}-\frac{\beta}{2}(s-1)(\nabla_{\mu_1}\nabla^{\alpha}
+\nabla^{\alpha}\nabla_{\mu_1})c_{\alpha\mu_2\ldots\mu_{s-1}}
\nonumber\\
&& + \frac{1}{4}(\beta+2)(s-1)^2 R c_{\mu_1\mu_2\ldots
\mu_{s-1}}\Big).\qquad
\end{eqnarray}

\noindent Taking into account that the ghost fields are totally
symmetric and traceless, it is easy to see that the matrices
needed for calculations based on Eq. (\ref{B2}) have the following
form:

\begin{eqnarray}\label{Ghost_Matrixes1}
&& \varepsilon
K^{\mu\nu}{}_{\alpha_1\ldots\alpha_{s-1}}{}^{\beta_1\ldots\beta_{s-1}}
= -\frac{1}{2}\beta(s-1)\cdot
P_{\alpha_1\ldots\alpha_{s-1}}{}^{\gamma_1\ldots\gamma_{s-1}}\nonumber\\
&&\qquad\qquad\qquad\qquad\qquad \times \Big(
g^{\mu (\delta_1} \delta^\nu_{(\gamma_1}
1^{\delta_2\ldots\delta_{s-1})}_{\gamma_2\ldots\gamma_{s-1})} +
g^{\nu (\delta_1} \delta^\mu_{(\gamma_1}
1^{\delta_2\ldots\delta_{s-1})}_{\gamma_2\ldots\gamma_{s-1})}
\Big) P_{\delta_1\ldots\delta_{s-1}}{}^{\beta_1\ldots\beta_{s-1}};\qquad\\
\label{Ghost_Matrixes2}
&& W_{\alpha_1\ldots\alpha_{s-1}}{}^{\beta_1\ldots\beta_{s-1}} =
\frac{1}{4} (\beta+1) (s-1)^2 R \cdot
P_{\alpha_1\ldots\alpha_{s-1}}{}^{\beta_1\ldots\beta_{s-1}},\vphantom{\Big(}
\end{eqnarray}

\noindent where
$P_{\alpha_1\ldots\alpha_{s-1}}{}^{\beta_1\ldots\beta_{s-1}}$ is a
projection operator to the traceless states in two dimensions. Its
structure and also some its properties are described in the Appendix.

Substituting the matrices (\ref{Ghost_Matrixes1}) and
(\ref{Ghost_Matrixes2}) into Eq. (\ref{B2}), after some simple
transformations we obtain that for diagrams with a loop of the
Faddeev--Popov ghosts the expression for $b_2$ coefficient is

\begin{equation}
b_{2(\mbox{\scriptsize gh})}= \Big((s-1)^2+\frac{1}{3}+
\frac{s(s-1)^2}{2} \beta +o(\beta)\Big)R.
\end{equation}

Let us sum the results for the main contribution and the
contribution of the Faddeev--Popov ghosts. Doing this, it is
necessary to take into account that the ghost fields are
anticommuting that gives the factor $(-1)$ and differ from
antighosts that gives the factor $2$. Therefore, the final
result for the divergent part of the one-loop effective action
for $s\ge 3$ is written as

\begin{equation}\label{Result}
\Gamma_{1-loop}^{(\infty)} = \frac{1}{4\pi(d-2)}\int
d^2x\sqrt{-g}\,\Big(b_{2(\mbox{\scriptsize main})}
-2b_{2(\mbox{\scriptsize gh})}\Big) = \frac{1}{4\pi(d-2)}\int
d^2x\sqrt{-g} \Big(-2 R + o(\lambda,\beta)\Big).
\end{equation}

\noindent This expression does not contain terms of the first
order in $\lambda$ and $\beta$. Thus, in the considered
approximation the result is gauge independent. Moreover, the
result does not depend on a value of $s$. For $s=3$ all equations
obtained here agree with the ones obtained in Ref. \cite{Popova:2012ipa},
in which this particular case has been considered.

\section{Conclusion} \hspace{\parindent}

In this paper a simple algorithm for calculating one-loop divergences
in two dimensions is proposed in the case when the second variation
of the action is a nonminimal operator of the second order, and
``nonminimal'' terms are small. It is important that the proposed
formula allows to calculate terms which are total derivatives. This
formula in the considered limit appeared to be very simple. It is
manifestly covariant and allows to make calculations easily on the
curved space background.

As an application we calculated one-loop divergences for the
higher spin theory on the constant curvature background in a
nonminimal gauge, which depends on the two parameters $\lambda$
and $\beta$, in the limit in which these parameters are small.
By an explicit calculation we demonstrated that in the considered
approximation the result is gauge independent. This follows from
the fact that the considered effective action is the Green functions
generating functional without sources, which does not depend on gauge.
Moreover, the calculations showed that the result is independent of
the spin value $s$ for $s\ge 3$. Vanishing of the gauge dependence
can be also considered as a test of Eq. (\ref{B2}) correctness, especially if
one takes into account that all intermediate expressions depend on
the gauge parameters in a highly nontrivial way.

Although the case $d=2$ is not so interesting as the case $d=4$,
the method used in this paper can be applied for making
calculations in other dimensions. In the considered limit (when
nonminimal terms are small) it is reasonable to expect that the result will
be much simpler than the general formula presented in Ref.
\cite{Pronin:1996rv}. Moreover, it becomes possible to take into account
total derivatives, which were omitted in Ref. \cite{Pronin:1996rv}.

Possibly, one can also try to find the answer for an arbitrary nonminimal
operator for which nonminimal terms are not small taking into
account the total derivative terms. However, this problem has not
so far been solved.

\medskip

{\it Note:} After appearing the first version of this paper on the ArXiv we learned about some 
results related to the ones obtained in this paper. In particular, the $b_2$ coefficient for 
the considered operator has been also calculated in \cite{Moss:2013cba} by a different method. 
We have verified that Eq. (\ref{B2}) is in agreement with this result. Moreover, the independence 
of the one-loop divergences on $s$ is possibly related to the cancellation of the vacuum energy 
in the sum over all spins, regularized by the help of zeta-function \cite{Giombi:2013yva}. This 
follows from the triviality of the partition function \cite{Beccaria:2015vaa}, in which contributions
of each spin is given by the ratio of functional determinants \cite{Gupta:2012he} that cancel
each other in the product.

\section*{Acknowledgments}
\hspace{\parindent}

The  work of K.S. is supported by the RFBR grant 14-01-00695. The authors are very grateful to
I.L.Buchbinder, S.M.Kuzenko, P.I.Pronin, A.A.Reshetnyak, and A.A.Tseytlin for
valuable comments and discussions.

\appendix

\section{Appendix: Projection operators to the traceless and double
traceless states and their properties.}
\hspace{\parindent}\label{Appendix_Projectors}

Since the higher spin fields are double traceless, and the
corresponding ghost fields are traceless, expressions for the
second variation of the action (or the ghost action) contain
projection operators to the double traceless (or traceless)
states. In this appendix we describe structure of these projection
operators and point out some their properties.

In the explicit form the projection operator to the traceless
states in two dimensions is written as

\begin{equation}
P_{\alpha_1\ldots\alpha_{s-1}}^{\beta_1\ldots\beta_{s-1}} =
1_{\alpha_1\ldots\alpha_{s-1}}^{\beta_1\ldots\beta_{s-1}} - x_1
\cdot g_{(\alpha_1\alpha_2} g^{(\beta_1\beta_2}
1_{\alpha_3\ldots\alpha_{s-1})}^{\beta_3\ldots\beta_{s-1})} -x_2
\cdot g_{(\alpha_1\alpha_2} g_{\alpha_3\alpha_4}
g^{(\beta_1\beta_2} g^{\beta_3\beta_4}
1_{\alpha_5\ldots\alpha_{s-1})}^{\beta_5\ldots\beta_{s-1})}
-\ldots,
\end{equation}

\noindent where $x_1$, $x_2$, $\ldots$ are numerical coefficients
depending on $s$ and $d$. In this paper we need not explicit
expressions for these coefficients. They can be found, e.g., in
Ref. \cite{Takata}. The traceless projection operator has the
following properties:

\begin{eqnarray}
&& P_{\alpha_1\alpha_2\ldots\alpha_{s-1}}
{}^{\beta_1\beta_2\ldots\beta_{s-1}} g^{\alpha_1\alpha_2} =
0;\qquad P_{\alpha_1\alpha_2\ldots\alpha_{s-1}}
{}^{\beta_1\beta_2\ldots\beta_{s-1}} g_{\beta_1\beta_2}
= 0;\vphantom{\Big(}\nonumber\\
&& P_{\alpha_1\alpha_2\ldots\alpha_{s-1}}
{}^{\beta_1\beta_2\ldots\beta_{s-1}}
P_{\beta_1\beta_2\ldots\beta_{s-1}}
{}^{\gamma_1\gamma_2\ldots\gamma_{s-1}} =
P_{\alpha_1\alpha_2\ldots\alpha_{s-1}}
{}^{\gamma_1\gamma_2\ldots\gamma_{s-1}}.\vphantom{\Big(}
\end{eqnarray}

\noindent Two first equalities are actually  a part of the projection
operator definition, and the last one is their straightforward consequence.
Moreover, in the case $d=2$ and $s\ge 2$

\begin{equation}
\mbox{tr}\, P =
P_{\alpha_1\ldots\alpha_{s-1}}{}^{\alpha_1\ldots\alpha_{s-1}} =2.
\end{equation}

\noindent This can be easily verified by calculating a number of
independent components for the traceless field in two dimensions.

Similarly, the projection operator to the double traceless states
has the form

\begin{eqnarray}
&& Q_{\alpha_1\alpha_2\ldots\alpha_s}
{}^{\beta_1\beta_2\ldots\beta_s} =
1_{\alpha_1\alpha_2\ldots\alpha_s}^{\beta_1\beta_2\ldots\beta_s} -
y_1 \cdot g_{(\alpha_1\alpha_2} g_{\alpha_3\alpha_4}
g^{(\beta_1\beta_2} g^{\beta_3\beta_4}
1_{\alpha_5\ldots\alpha_s)}^{\beta_5\ldots\beta_s)}\vphantom{\Big(}\nonumber\\
&& - y_2 \cdot g_{(\alpha_1\alpha_2} g_{\alpha_3\alpha_4}
g_{\alpha_5\alpha_6} g^{(\beta_1\beta_2} g^{\beta_3\beta_4}
g^{\beta_5\beta_6}
1_{\alpha_7\ldots\alpha_s)}^{\beta_7\ldots\beta_s)}\vphantom{\Big(}
-\ldots,
\end{eqnarray}

\noindent where $y_1$, $y_2$, $\ldots$ are numerical coefficients
depending on $s$ and $d$, which satisfy the following properties:

\begin{eqnarray}
&& Q_{\alpha_1\alpha_2\ldots\alpha_s}
{}^{\beta_1\beta_2\ldots\beta_s} g^{\alpha_1\alpha_2}
g^{\alpha_3\alpha_4} = 0;\qquad Q_{\alpha_1\alpha_2\ldots\alpha_s}
{}^{\beta_1\beta_2\ldots\beta_s} g_{\beta_1\beta_2}
g_{\beta_3\beta_4}
= 0;\vphantom{\Big(}\nonumber\\
&& Q_{\alpha_1\alpha_2\ldots\alpha_s}
{}^{\beta_1\beta_2\ldots\beta_s} Q_{\beta_1\beta_2\ldots\beta_s}
{}^{\gamma_1\gamma_2\ldots\gamma_s} =
Q_{\alpha_1\alpha_2\ldots\alpha_s}
{}^{\gamma_1\gamma_2\ldots\gamma_s}.\vphantom{\Big(}\nonumber\\
\end{eqnarray}

\noindent In the case $d=2$ the following identities are also valid:

\begin{eqnarray}
&& \mbox{tr}\, Q = Q_{\alpha_1\alpha_2\ldots\alpha_s}
{}^{\alpha_1\alpha_2\ldots\alpha_s} = 4, \qquad\mbox{if}\qquad s\ge 3;\vphantom{\Big(}\nonumber\\
&& Q_{\alpha_1\alpha_2\ldots\alpha_s}
{}^{\beta_1\beta_2\ldots\beta_s} g^{\alpha_1\alpha_2}
g_{\beta_1\beta_2} = \frac{4}{s} \cdot
P_{\alpha_3\ldots\alpha_s}{}^{\beta_3\ldots\beta_s}.\vphantom{\Big(}
\end{eqnarray}

\noindent The trace of the projection operator $Q$ can be found by
calculating a number of independent components for the
double traceless field in two dimensions. In order to verify the
last identity we note that the left hand side is evidently
proportional to the traceless projection operator, and the
coefficient can be found by comparing terms proportional to
$1_{\alpha_3\ldots\alpha_s}^{\beta_3\ldots\beta_s}$.



\begin{thebibliography}{100}

\bibitem{'tHooft:1974bx}
  G.~'t Hooft and M.~J.~G.~Veltman,
  Annales Poincare Phys.\ Theor.\ A {\bf 20} (1974) 69.

\bibitem{Pronin:1996rv}
  P.~I.~Pronin and K.~V.~Stepanyantz,
  Nucl.\ Phys.\ B {\bf 485} (1997) 517.

\bibitem{Kazakov:1997nq}
  K.~A.~Kazakov, P.~I.~Pronin and K.~V.~Stepanyantz,
  Grav.\ Cosmol.\  {\bf 4} (1998) 17.

\bibitem{Kalmykov:1998cv}
  M.~Y.~Kalmykov, K.~A.~Kazakov, P.~I.~Pronin and K.~V.~Stepanyantz,
  Class.\ Quant.\ Grav.\  {\bf 15} (1998) 3777.

\bibitem{Kazakov:1998qt}
  K.~A.~Kazakov and P.~I.~Pronin,
  Phys.\ Rev.\ D {\bf 59} (1999) 064012.

\bibitem{Kazakov:1999qd}
  K.~A.~Kazakov and P.~I.~Pronin,
  Theor.\ Math.\ Phys.\  {\bf 121} (1999) 1585
  [Teor.\ Mat.\ Fiz.\  {\bf 121} (1999) 387].

\bibitem{Andriyash:2002yb}
  E.~Andriyash and K.~Stepanyantz,
  Moscow Univ.\ Phys.\ Bull.\  {\bf 58} (2003) 28
  [Vestn.\ Mosk.\ Univ.\ Fiz.\ Astron.\  {\bf 58} (2003) 24].

\bibitem{Bolotin:2013qgr}
  D.~A.~Bolotin and S.~V.~Poslavsky,
  ``Introduction to Redberry: the computer algebra system designed for tensor manipulation,''
  arXiv:1302.1219 [cs.SC].  

\bibitem{Poslavsky:2015vba}
  S.~Poslavsky and D.~Bolotin,
  J.\ Phys.\ Conf.\ Ser.\  {\bf 608} (2015) 1,  012060.

\bibitem{Minakshisundaram:1949xg}
  S.~Minakshisundaram and A.~Pleijel,
  Can.\ J.\ Math.\  {\bf 1} (1949) 242.

\bibitem{Minakshisundaram:1953xh}
  S.~Minakshisundaram,
  J.\ Indian Math.\ Soc.\  {\bf 17} (1953) 158.

\bibitem{Seeley:1967ea}
  R.~T.~Seeley,
  Proc.\ Symp.\ Pure Math.\  {\bf 10} (1967) 288.

\bibitem{Gilkey:1975iq}
  P.~B.~Gilkey,
  J.\ Diff.\ Geom.\  {\bf 10} (1975) 601.

\bibitem{Schwinger:1951nm}
  J.~S.~Schwinger,
  Phys.\ Rev.\  {\bf 82} (1951) 664.

\bibitem{DeWitt:1965jb}
  B.~S.~DeWitt,
  ``Dynamical theory of groups and fields,''
  Conf.\ Proc.\ C {\bf 630701} (1964) 585
  [Les Houches Lect.\ Notes {\bf 13} (1964) 585].

\bibitem{Barvinsky:1985an}
  A.~O.~Barvinsky and G.~A.~Vilkovisky,
  Phys.\ Rept.\  {\bf 119} (1985) 1.

\bibitem{Gusynin:1990ek}
  V.~P.~Gusynin, E.~V.~Gorbar and V.~V.~Romankov,
  Nucl.\ Phys.\ B {\bf 362} (1991) 449.

\bibitem{Camporesi:1990wm}
  R.~Camporesi,
  Phys.\ Rept.\  {\bf 196} (1990) 1.

\bibitem{Camporesi:1993mz}
  R.~Camporesi and A.~Higuchi,
  Phys.\ Rev.\ D {\bf 47} (1993) 3339.

\bibitem{Takata}
I.~L.~Buchbinder, V.~A.~Krykhtin and H.~Takata, ``One-loop
divergence of effective action of Higher spin theory in
$\mbox{AdS}_4$,'' a talk given at the conference QFTG-2010.

\bibitem{Huang:1982ik}
  K.~Huang,
  ``Quarks, Leptons And Gauge Fields,''  Singapore, Singapore: World Scientific (1992) 333p.

\bibitem{Pronin:1997eb}
  P.~I.~Pronin and K.~Stepanyantz,
  Phys.\ Lett.\ B {\bf 414} (1997) 117.

\bibitem{'tHooft:1972fi}
  G.~'t Hooft and M.~J.~G.~Veltman,
  Nucl.\ Phys.\ B {\bf 44} (1972) 189.

\bibitem{Bollini:1972ui}
  C.~G.~Bollini and J.~J.~Giambiagi,
  Nuovo Cim.\ B {\bf 12} (1972) 20.

\bibitem{Ashmore:1972uj}
  J.~F.~Ashmore,
  Lett.\ Nuovo Cim.\  {\bf 4} (1972) 289.

\bibitem{Cicuta:1972jf}
  G.~M.~Cicuta and E.~Montaldi,
  Lett.\ Nuovo Cim.\  {\bf 4} (1972) 329.

\bibitem{Obukhov:1983mm}
  Y.~N.~Obukhov,
  Nucl.\ Phys.\ B {\bf 212} (1983) 237.

\bibitem{Fronsdal:1978rb}
  C.~Fronsdal,
  Phys.\ Rev.\ D {\bf 18} (1978) 3624.

\bibitem{Fang:1978wz}
  J.~Fang and C.~Fronsdal,
  Phys.\ Rev.\ D {\bf 18} (1978) 3630.

\bibitem{Vasiliev:2004qz}
  M.~A.~Vasiliev,
  Fortsch.\ Phys.\  {\bf 52} (2004) 702.

\bibitem{Bekaert:2005vh}
  X.~Bekaert, S.~Cnockaert, C.~Iazeolla and M.~A.~Vasiliev,
  ``Nonlinear higher spin theories in various dimensions,''
  hep-th/0503128.

\bibitem{Fronsdal:1978vb}
  C.~Fronsdal,
  Phys.\ Rev.\ D {\bf 20} (1979) 848.

\bibitem{Buchbinder:2011vp}
  I.~L.~Buchbinder, V.~A.~Krykhtin and P.~M.~Lavrov,
  Mod.\ Phys.\ Lett.\ A {\bf 26} (2011) 1183.

\bibitem{Faddeev:1980be}
  L.~D.~Faddeev and A.~A.~Slavnov,
``Gauge Fields. Introduction To Quantum Theory,'' Nauka, Moscow,
1978 and
    [Front.\ Phys.\  {\bf 83} (1990) 1]
    [Benjamin, Reading, 1990].

\bibitem{Faddeev:1967fc}
  L.~D.~Faddeev and V.~N.~Popov,
  Phys.\ Lett.\ B {\bf 25} (1967) 29.

\bibitem{Kallosh:1978wt}
  R.~E.~Kallosh, O.~V.~Tarasov and I.~V.~Tyutin,
  Nucl.\ Phys.\ B {\bf 137} (1978) 145.

\bibitem{Popova:2012ipa}
  H.~Popova and K.~Stepanyantz,
  TSPU Bulletin {\bf 2012} (2012) 13,  130.

\bibitem{Moss:2013cba}
  I.~G.~Moss and D.~J.~Toms,
  J.\ Phys.\ A {\bf 47} (2014) 215401.

\bibitem{Giombi:2013yva}
  S.~Giombi, I.~R.~Klebanov, S.~S.~Pufu, B.~R.~Safdi and G.~Tarnopolsky,
  JHEP {\bf 1310} (2013) 016.

\bibitem{Beccaria:2015vaa}
  M.~Beccaria and A.~A.~Tseytlin,
  J.\ Phys.\ A {\bf 48} (2015) 27,  275401.

\bibitem{Gupta:2012he}
  R.~K.~Gupta and S.~Lal,
  JHEP {\bf 1207} (2012) 071.

\end{thebibliography}
\end{document}